# Single Mode Approximation for sub-Ohmic Spin-Boson Model: Adiabatic Limit and Critical Properties

Fei-Ran Liu[1] and Ning-Hua Tong[1,a]

Department of Physics, Renmin University of China, Beijing 100872, P. R. China

**Abstract.** In this work, the quantum phase transition in the sub-Ohmic spin-boson model is studied using a single-mode approximation, by combining the rotating wave transformation and the transformations used in the numerical renormalization group (NRG). Analytical results for the critical coupling strength $\alpha_c$, the magnetic susceptibility $\chi(T)$, and the spin-spin correlation function $C(\omega)$ at finite temperatures are obtained and further confirmed by numerical results. We obtain the same $\alpha_c$ as the mean-field approximation. The critical exponents are classical: $\beta = 1/2$, $\delta = 3$, $\gamma = 1$, $x = 1/2$, $y_t^* = 1/2$, in agreement with the spin-boson model in $0 < s < 1/2$ regime. $C(\omega)$ has nontrivial behavior reflecting coherent oscillation with temperature dependent damping effects due to the environment. We point out the original NRG has problem with the crossover temperature $T^*$, and propose a chain Hamiltonian possibly suitable for implementing NRG without boson state truncation error.

## 1 Introduction

The spin-boson model (SBM) is the simplest model that describes a quantum two-level system coupled to a dissipative environment. SBM has extensive relevance to physical systems in condensed matter physics [1,2], quantum optics, and quantum chemistry. Its properties are studied extensively. For the bath spectra exponent $0 \leq s < 1$ (sub-ohmic bath), the ground state may change from the spin-tunneling state to the spin-pinned state through a second order phase transition, as the dissipation strength increase above a critical value. This environment-induced quantum phase transition attracts much attention in the past few years [3,4,5,6,7,8,9,11,12] and the universality class of this transition is under intensive studies [13,14,15,16,17,18,19,20,21,22,23].

The standard quantum-classical mapping predicts that the sub-Ohmic SBM is equivalent to a classical Ising spin chain with long-range interaction [1]. Classical Monte Carlo simulations [24] and renormalization group analysis [25] for the long-range Ising model predict a continuous magnetic transition with classical critical exponents for $0 < s < 1/2$, i.e., $\beta = 1/2$, $\delta = 3$, and $\gamma = 1$. In Ref. [4], using the numerical renormalization group (NRG) technique, a continuous transition was found in the sub-Ohmic SBM, with non-classical exponents for $0 < s < 1/2$. Recently, other methods such as quantum monte carlo (QMC) [16], sparse polynomial space approach [8], and extended coherent state techniques [6,7] have found classical critical exponents for $0 < s < 1/2$.

Later research disclosed that although non-universal quantities such as $\alpha_c$ are obtained quite accurately, NRG results are not reliable for critical exponents $\beta$, $\delta$, and $x$ in the deep sub-Ohmic regime $0 < s < 1/2$. This is due to two inherent limitations of the bosonic NRG (BNRG): the boson Hilbert-space truncation error and the mass flow error [15]. On one hand, the BNRG employs boson Hilbert-space truncation in the iterative diagonalization process. It causes errors in $\beta$ and $\delta$ which characterize the flow into the localized phase at zero temperature [15,23]. On the other hand, the mass flow causes errors for non-zero temperatures as a result of neglecting of low-lying bath modes with energy smaller than temperature [18,19].

Recently, in Ref. [23], the BNRG calculation is combined with finite size scaling analysis of $N_b$, the number of truncated boson states, to recover the correct exponents $\beta$ and $\delta$. Using the matrix product state to describe the ground state of the Wilson chain, Guo *et al.* were able to take the optimal boson basis and successfully overcome the boson state truncation error. However, for accurate study of finite temperature as well as dynamical properties, a NRG-like algorithm without boson state truncation is still desirable, with its energy flow incorporating physical information of excited states. As an attempt to find possible NRG variants for this purpose, we carry out the rotating wave transformation (RWT) [10,11,12,26] to the Hamiltonian of the SBM, to single out the boson displacement induced by the spin-boson coupling, so that the truncation errors for the displaced boson modes could be avoided in the NRG calculation. In the limit of complete transformation $\eta \to 0$ ($\eta$ is a control parameter of RWT), a boson mode is singled out in the Wilson chain Hamiltonian, which has vanishing frequency and hopping amplitude to other sites. In this paper, we consider the single mode Hamiltonian $H_0(\eta)$ in which the spin is coupled only to this adiabatic mode, as the first step towards the BNRG study of the full chain. Using adiabatic approximation which is exact in the adiabatic limit $\eta \rightarrow 0$ and checking it using exact diagonalization method, we find that this model gives the critical coupling strength $\alpha_c = s\Delta/2\omega_c$, which is same as the mean-field result [23]. The critical exponents are obtained as $\beta = 1/2$, $\delta = 3$, $\gamma = 1$, $x = 1/2$, and $y^*_t = 1/2$, in agreement with the expected classical exponents of the SBM in the regime $0 < s < 1/2$.

---

[a] e-mail: nhtong@ruc.edu.cn

The remainder of the paper is organized as follows: In Sec.2, we introduce the SBM and derive the approximate single-mode Hamiltonian $H_0(\eta)$ for it. In Section 3, we present results for the order parameter $\langle \sigma_z \rangle$, susceptibility $\chi(T)$ and spin-spin correlation function $C(\omega)$. Related issues including the scaling relations, the BNRG result $T^*$, and a possible NRG algorithm without boson state truncation error are discussed. Finally, we close with a summary.

## 2 Model and Method

### 2.1 effective single mode Hamiltonian $H_0(\eta)$

The Hamiltonian of SBM reads($\hbar = 1$) [1,2]

$$H_{SB} = -\frac{\Delta}{2}\sigma_x + \frac{\epsilon}{2}\sigma_z + \sum_i \omega_i a_i^\dagger a_i + \frac{\sigma_z}{2}\sum_i \lambda_i(a_i^\dagger + a_i). \quad (1)$$

Here, the two-state system is represented by the Pauli matrices $\sigma_x$ and $\sigma_z$. $\epsilon$ is the bias of the two states and $\Delta$ is the tunneling strength between them. The environment is modeled by a collection of harmonic oscillators, with creation and annihilation operators $a_i^\dagger$ and $a_i$, respectively. $\omega_i$ and $\lambda_i$ represent the frequency of the $i$-th boson mode and its coupling to the two-state system, respectively. The effect of the harmonic environment is characterized by the bath spectral function

$$J(\omega) \equiv \pi \sum_i \lambda_i^2 \delta(\omega - \omega_i). \quad (2)$$

We use a model spectral function as

$$J(\omega) = 2\pi\alpha\omega_c^{1-s}\omega^s, \quad (0 < \omega < \omega_c, \ s > -1), \quad (3)$$

where the dimensionless parameter $\alpha$ characterizes the dissipation strength, and the cutoff $\omega_c = 1$ is set as the energy unit.

Starting from Eq.(1), we carry out RWT [10,11,12,26] $\tilde{H} = UH_{SB}U^{-1}$. Here $U = e^{\hat{S}}$ and $\hat{S} = 1/2 \sum_k g_k(a_k^\dagger - a_k)\sigma_z$. $g_k$ is the parameter controlling the transformation for the $k$-th boson mode, for which we use

$$g_k = \frac{\lambda_k}{\omega_k + \eta}. \quad (4)$$

Here $\eta$ is the parameter to control the RWT. $\tilde{H}$ is then written as (neglecting a constant)

$$\tilde{H} = \frac{\epsilon}{2}\sigma_z + \sum_k \omega_k a_k^\dagger a_k + \frac{\eta}{2}\sigma_z \sum_k g_k(a_k^\dagger + a_k)$$
$$- \frac{\Delta}{2}\sigma_x \cosh(\hat{\chi}) - \frac{i}{2}\Delta\sigma_y \sinh(\hat{\chi}), \quad (5)$$

where $\hat{\chi} = \sum_k g_k\left(a_k^\dagger - a_k\right)$.

Similar RWT's have been widely used in the study of SBM. Combined with perturbation [11,26] or numerical diagonalization [6], both critical coupling strength $\alpha_c$ and dynamical properties are accurately obtained. $\eta$ measures how much displacement in the $k$-th mode induced by coupling to $\sigma_z$ is absorbed into the new boson mode. For $\eta = 0$, the coupling term $\sigma_z\left(a_k^\dagger + a_k\right)$ disappears in $\tilde{H}$. This term is the source of the displacement divergence in low energy boson modes, when the coupling is strong. It is expected that in the limit $\eta \rightarrow 0$, RWT removes most of the boson displacement and it thus helps to avoid the boson state truncation error in BNRG. Indeed, for the case $\Delta = 0$, $\tilde{H}(\eta = 0)$ has no displaced bosons.

For $\tilde{H}$, we define the bath spectral function as

$$\tilde{J}(\omega) \equiv \pi \sum_k (\eta g_k)^2 \delta(\omega - \omega_k)$$
$$= \frac{\eta^2}{(\omega + \eta)^2}J(\omega). \quad (6)$$

Note that in Eq.(5), the local boson modes that are coupled to $\sigma_z$ and to $\sigma_x$ are the same, i.e., $\sum_k g_k\hat{a}_k$. This is a result of the special form of $g_k$ in Eq.(4). With possible NRG-related applications in mind, we further carry out logarithmic discretization for Eq.(5), following the standard procedure [4]. We obtain the star-Hamiltonian as

$$\tilde{H}_s = \frac{\epsilon}{2}\sigma_z + \sum_{n=0}^{+\infty}\xi_n a_n^\dagger a_n + \frac{1}{2\sqrt{\pi}}\sigma_z\sum_{n=0}^{+\infty}\gamma_n(a_n + a_n^\dagger)$$
$$- \frac{\Delta}{2}\sigma_x \cosh(\hat{\chi}) - \frac{i\Delta}{2}\sigma_y \sinh(\hat{\chi}), \quad (7)$$

with

$$\xi_n = \frac{1}{\gamma_n^2}\int_{\Lambda^{-(n+1)}\omega_c}^{\Lambda^{-n}\omega_c}\tilde{J}(\omega)\omega d\omega$$
$$\gamma_n^2 = \int_{\Lambda^{-(n+1)}\omega_c}^{\Lambda^{-n}\omega_c}\tilde{J}(\omega)d\omega. \quad (8)$$

In Eq.(7), $\hat{\chi} = 1/(\eta\sqrt{\pi})\sum_{n=0}^{+\infty}\gamma_n(a_n^\dagger - a_n)$. $\Lambda > 1$ is the logarithmic parameter.

In Fig.1, we plot $\tilde{J}(\omega)$ on log scale and the coefficients $\xi_n$ and $\gamma_n/\eta$ for various $\eta$'s at $s = 0.3$. $\tilde{J}(\omega)$ is characterized by a crossover scale $\omega^* \sim \eta$, which separates $\tilde{J}(\omega) \sim \omega^s$ in $\omega << \omega^*$ and $\tilde{J}(\omega) \sim \omega^{s-2}$ in $\omega >> \omega^*$ regime. Correspondingly, $\gamma_n \sim \Lambda^{(n/2)(1-s)}$ in small $n$ and $\gamma_n \sim \Lambda^{-(n/2)(1+s)}$ in large $n$ regime. $\xi_n$ is independent of $\eta$ in the limit $\eta \rightarrow 0$. In the limit $\eta \rightarrow \infty$, $\tilde{J}(\omega) = J(\omega)$ and $\tilde{H}_s$ recovers the original star-Hamiltonian of SBM [4]. In the limit $\eta \rightarrow 0$, $\tilde{J}(\omega) \rightarrow 0$. $\gamma_n$ disappears but $\gamma_n/\eta$ follows $\Lambda^{(n/2)(1-s)}$ and diverges in large $n$ limit. As a result, the coupling term $\gamma_n\sigma_z(a_n + a_n^\dagger)$ in $\tilde{H}_s$ will disappear as $\eta \rightarrow 0$. The role of coupling operators are now played by $\cosh(\hat{\chi})$ and $\sinh(\hat{\chi})$ in which $\hat{\chi}$ is a local boson mode with diverging coefficient.

Using orthogonal transformation, the star-Hamiltonian can be transformed into a semi-infinite chain which is suitable for iterative diagonalization in NRG [29].

$$\tilde{H}_c = \frac{\epsilon}{2}\sigma_z + \frac{1}{2}\sqrt{\frac{\eta_0}{\pi}}\sigma_z\left(b_0 + b_0^\dagger\right) - \frac{\Delta}{2}\sigma_x \cosh(\hat{\chi})$$

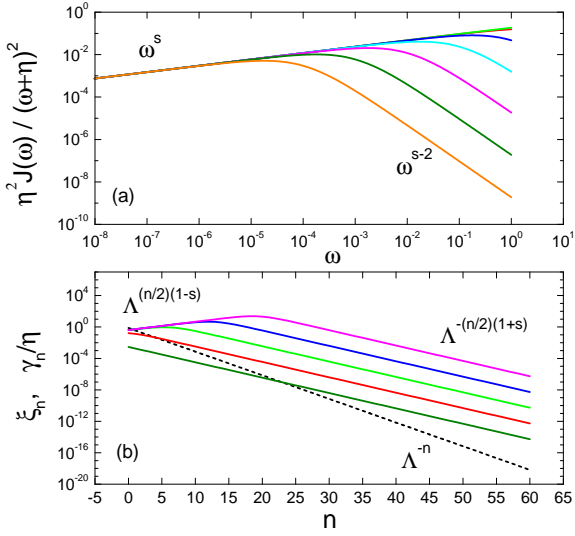

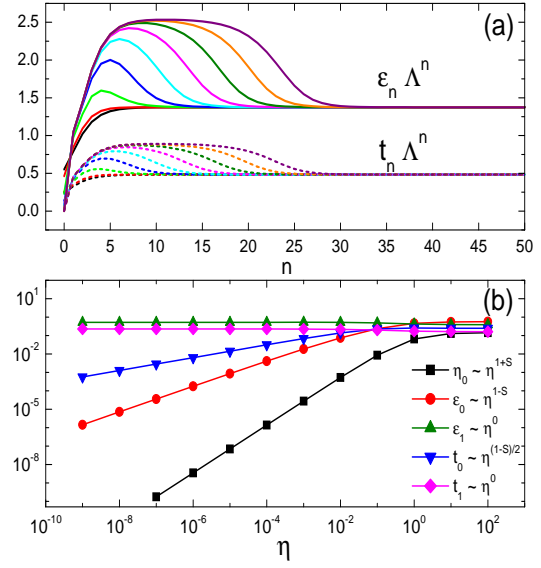

**Fig. 1.** (a) Transformed boson spectral function $\tilde{J}(\omega)$. From top to bottom: $\eta = 10^2, 10^1, 10^0, 10^{-1}, 10^{-2}, 10^{-3}, 10^{-4}$. (b) $\xi_n$ (dashed line independent of $\eta$) and $\gamma_n/\eta$ (solid lines). From bottom to top: $\eta = 10^2, 10^0, 10^{-2}, 10^{-4}, 10^{-6}$. Other parameters are $S = 0.3, \Delta = 0.1, \alpha = 0.03$, and $\Lambda = 2.0$

**Fig. 2.** (a) Chain coefficients $\epsilon_n \Lambda^n$ (solid lines) and $t_n \Lambda^n$ (dashed lines) as functions of $n$ for various $\eta$'s. From bottom to top, $\eta = 10^1, 10^0, ..., 10^{-7}$. (b) Some coefficients as functions of $\eta$: $\eta_0$ (squares), $\epsilon_0$ (circles), $\epsilon_1$ (up triangles), $t_0$ (down triangles), and $t_1$ (diamonds). Other parameters are $S = 0.3, \Delta = 0.1, \alpha = 0.03$, and $\Lambda = 2.0$

$$-\frac{\Delta}{2}i\sigma_y \sinh(\hat{\chi}) + \sum_{n=0}^{+\infty}\left[\varepsilon_n b_n^\dagger b_n + t_n(b_n^\dagger b_{n+1} + b_{n+1}^\dagger b_n)\right]. \quad (9)$$

with $\hat{\chi} = 1/\eta \sqrt{\eta_0/\pi}\left(b_0^\dagger - b_0\right)$. The parameters $\eta_0 = \sum_n \gamma_n^2$. $\varepsilon_n$ and $t_n$'s are obtained through the recursion relations as [4]

$$t_m = \left[\sum_{n=0}^{+\infty}[(\xi_n - \varepsilon_m)U_{mn} - t_{m-1}U_{m-1n}]^2\right]^{\frac{1}{2}},$$

$$\varepsilon_m = \sum_{n=0}^{+\infty} \xi_n U_{mn}^2,$$

$$U_{m+1n} = \frac{1}{t_m}\left[(\xi_n - \varepsilon_m)U_{mn} - t_{m-1}U_{m-1n}\right]. \quad (10)$$

The initial conditions are $t_{-1} = 0$, $U_{-1n} = 0$, and $U_{0n} = \gamma_n/\sqrt{\eta_0}$.

Now we analyze the asymptotic form of the coefficients in $\tilde{H}_c$. The coupling constant behaves as $\eta_0 \propto \eta^{1+s}$ so that the term $\sigma_z\left(b_0^\dagger + b_0\right)$ disappears in the limit $\eta \to 0$. The strong coupling fixed point at $\Delta = 0$ can then be obtained correctly because no displacements of bosons are present. The price is that the parameter in $\hat{\chi}$ diverges as $\sqrt{\eta_0}/\eta \propto \eta^{(-1+s)/2}$. In Fig.2(a), we plot $\epsilon_n \Lambda^n$ and $t_n \Lambda^n$ as functions of $n$, for various $\eta$'s at $s = 0.3$. It is seen that $\epsilon_n$ and $t_n$ have a plateau in high energy regime $n < n_{cr}$. While for $n > n_{cr}$, they approach the levels same as in the original NRG chain. Here, $n_{cr}$ is a crossover scale related to the $\omega^*$ of $\tilde{J}(\omega)$. The plateaus in the curves extend to larger $n$ for smaller $\eta$. In the limit $\eta \to 0$, they form new levels. This means that $\epsilon_n(\eta = 0)$ and $t_n(\eta = 0)$ have different values from $\epsilon_n(\eta = \infty)$ and $t_n(\eta = \infty)$ but still decays as $\Lambda^{-n}$. This is a necessary condition for NRG to be applicable for $\tilde{H}_c(\eta = 0)$.

Direct NRG calculation based on $\tilde{H}_c$ shows that although the original BNRG results is recovered using $\eta \geq 100$, the flow lines fails to describe a fixed point for $\eta \leq 10^{-2}$. The reason lies in the following asymptotic relations as shown in Fig.2(b),

$$\epsilon_0 \propto \eta^{1-s}, \quad t_0 \propto \eta^{(1-s)/2},$$
$$\epsilon_1 \propto \eta^0, \quad t_1 \propto \eta^0. \quad (11)$$

In the limit $\eta = 0$, $\epsilon_0 = t_0 = 0$. This means that the zero-th mode of the boson chain becomes an adiabatic boson mode with $\Delta/\epsilon_0 \to \infty$. At the same time, it tends to get disconnected to the rest of the chain due to $t_0 \to 0$. The Hamiltonian $\tilde{H}_0$ which contains the spin and the 0-th boson mode reads

$$\tilde{H}_0 = \frac{\varepsilon}{2}\sigma_z + \frac{1}{2}\sqrt{\frac{\eta_0}{\pi}}\sigma_z(b_0 + b_0^\dagger) + \varepsilon_0 b_0^\dagger b_0$$
$$- \frac{\Delta}{2}\sigma_x \cosh(\hat{\chi}) - \frac{\Delta}{2}i\sigma_y \sinh(\hat{\chi}). \quad (12)$$

When doing iterative diagonalization within NRG, $\tilde{H}_0$ is first diagonalized and the second boson mode $b_1$ is added to form a new Hamiltonian. In the limit $\epsilon = 0$, the energy levels of $\tilde{H}_0$ tends to be infinitely dense and is close to degeneracy. As a result, the next boson site to be added will not be a small perturbation to $\tilde{H}_0$. The basis for NRG to work, i.e., the exponential decay of energy scale, is not fulfilled.

Although the most direct endeavor of taking $\eta \to 0$ does not give a useful chain Hamiltonian for BNRG, the obtained single-mode Hamiltonian $\tilde{H}_0$ is still interesting. First, it represents the limit of long-range interaction in the imaginary time axis, and its critical properties should show the characteristics of the SBM in small $s$ limit. Second, $\tilde{H}_0$ describes a spin coupled to an adiabatic resonator which by itself is an interesting system. Previous study in terms of the entanglement entropy shows that a quantum phase transition occurs in this system [27]. Third, as an exactly solvable model, $\tilde{H}_0$ serves as the zeroth order approximation to $\tilde{H}_c$ in the sense that the full model can be studied by treating the hopping term $t_0\left(b_0^\dagger b_1 + b_1^\dagger b_0\right)$ as a small perturbation. Therefore, in the following sections, we will focus on the properties of $\tilde{H}_0(\eta \to 0)$ and leave the full Hamiltonian $\tilde{H}_c$ for a future study.

## 2.2 exact solution in the adiabatic limit

As an approximation to $\tilde{H}_c$, we truncate the chain at the 0-th site and obtain $\tilde{H}_0(\eta)$. It is noted that although $t_0$ tends to zero in the limit $\eta$ to 0, this truncation is not exact even in this limit. The reason is related to the singular coefficient in the operator $\hat{\chi}$ in $\tilde{H}_0$, to be discussed later. In order to study $\tilde{H}_0(\eta)$, we use the adiabatic approximation which is exact in the limit $\eta \to 0$. We check our results using the exact diagonalization method.

To do the adiabatic approximation, we carry out another RWT $\bar{H}_0 = V\tilde{H}_0 V^{-1}$ where $V = exp[\kappa\lambda(b^\dagger - b)]\sigma_z$. Choosing a suitable parameter $\kappa$, we obtain

$$\bar{H}_0 = \frac{\varepsilon}{2}\sigma_z - \frac{\Delta}{2}\sigma_x + \frac{1}{2}\lambda\sigma_z(b + b^\dagger) + \varepsilon_0 b^\dagger b, \quad (13)$$

where $\lambda = \sqrt{\eta_0/\pi}(1 + \varepsilon_0/\eta)$. From Eq.(11), one obtains $\lambda \sim \eta^{(1-s)/2}$ and the $\varepsilon_0 \sim \eta^{1-s}$. It is easy to show that $\Delta/\varepsilon_0 \to \infty$ and $\lambda/\varepsilon_0 \to \infty$, meaning that the limit of $\eta = 0$ is the adiabatic limit.

Using the replacement $q = (b^\dagger + b)/\sqrt{2}$ and $p = i(b^\dagger - b)/\sqrt{2}$, we get

$$\bar{H}_0 = H_a[q, \sigma_z, \sigma_x] + H_d[p], \quad (14)$$

with

$$H_a = \frac{\varepsilon}{2}\sigma_z - \frac{\Delta}{2}\sigma_x + \frac{1}{\sqrt{2}}\lambda\sigma_z q + \frac{1}{2}\varepsilon_0 q^2,$$
$$H_d = \frac{1}{2}\varepsilon_0 p^2. \quad (15)$$

In the adiabatic limit, the dynamical part $H_d$ does not play any role and the partition function of $\bar{H}_0$ and the average of a physical observable $\hat{O}$ can be evaluated exactly by

$$Z(\beta) = \int_{-\infty}^{\infty} dq Tr e^{-\beta H_a[q,\sigma_z,\sigma_x]},$$
$$\langle \hat{O} \rangle = \frac{1}{Z(\beta)} \int_{-\infty}^{\infty} dq Tr \left[\hat{O} e^{-\beta H_a[q,\sigma_z,\sigma_x]}\right]. \quad (16)$$

Here the trace is for the spin degrees of freedom [27]. After redefining $\sqrt{2}\lambda q \to q$, $H_a$ becomes

$$H_a = \frac{\varepsilon}{2}\sigma_z - \frac{\Delta}{2}\sigma_x + \frac{1}{2}\sigma_z q + \frac{c}{4}q^2. \quad (17)$$

Here $c = \varepsilon_0/\lambda^2$. The rescaling of $q$ changes a factor in $Z$ and does not influence the physical results. In the limit $\eta \to 0$, one can prove using Eq.(10)-(11) that $c = s/(2\alpha\omega_c)$ is a constant, independent of $\Lambda$.

For numerical calculations, we start from $\tilde{H}_0$ in Eq.(12). We build matrices for $\cosh(\hat{\chi})$ and $\sinh(\hat{\chi})$ on $N_b$ boson occupation bases $\{|0>, |1>, ..., |N_b - 1>\}$, using the eigen wave functions of a displaced harmonic oscillator [28]. In the adiabatic limit, the displacement is strong. We use an efficient algorithm to produce the associated Laguerre polynomial for calculating the wave functions. The details are presented in Appendix B.

## 3 Results and Discussions

### 3.1 magnetization $\langle \sigma_z \rangle$

For physical analysis, we calculate the average of $\hat{\sigma}_z$ at zero temperature, the static susceptibility $\chi(T)$ and the spin-spin correlation function $C(\omega)$ at finite temperatures. Here we present the main results and put the details of derivation in Appendix A.

The zero temperature magnetization for $\alpha > \alpha_c$ is determined by the following equations,

$$\langle \sigma_z \rangle = -cq_0, \quad (18)$$

with $q_0$ being one of the solutions of the equation below,

$$cq = \frac{\epsilon + q}{\sqrt{\Delta^2 + (\epsilon + q)^2}}. \quad (19)$$

From this expression, the critical coupling strength is obtained in the limit $\eta = 0$ as $\alpha_c = \Delta s/(2\omega_c)$, which is exactly the mean-field result [23]. Here, due to the error introduced in truncating $\tilde{H}_c$ to $\tilde{H}_0$, $\tilde{H}_0$ only qualitatively describe the effect of the bath. The critical behavior of $\langle \sigma_z \rangle$ can be identified exactly from Eq.(19) and it is also same as the mean-field result,

$$\langle \sigma_z \rangle (\epsilon = 0) \propto (\alpha - \alpha_c)^{1/2} \quad (\alpha > \alpha_c);$$
$$\langle \sigma_z \rangle (\alpha = \alpha_c) \propto \epsilon^{1/3}. \quad (20)$$

This shows that our single-mode approximation gives the classical critical exponent $\beta = 1/2$ and $\delta = 3$, which are the correct result for the SBM in the regime $0 < s < 1/2$.

In Fig.3, $m \equiv \langle \sigma_z \rangle/2$ as functions of $\epsilon$ are plotted on log scale for various $\alpha$ values at $s = 0.3$. For $\alpha < \alpha_c$, $m \propto \chi\epsilon$. The susceptibility $\chi \propto (\alpha - \alpha_c)^{-\gamma}$ when $\alpha$ approaches $\alpha_c$. For $\alpha = \alpha_c$, $m \propto \epsilon^{1/\delta}$. For $\alpha > \alpha_c$, $m \propto \epsilon^0(\alpha - \alpha_c)^\beta$ in the small $\epsilon$ limit. Numerical fit of data in Fig.3 confirms the classical exponents $\beta = 1/2$, $\delta = 3$, and $\gamma = 1$ (inset of Fig.3). We observe that the $m - \epsilon$ curves for $\alpha > \alpha_c$ and for $\alpha < \alpha_c$ always merge with the curve for $\alpha = \alpha_c$ in the regime $\epsilon > \epsilon_{cr}$. Here $\epsilon_{cr}$ is the crossover value separating the quantum critical regime and the delocalized phase (for $\alpha < \alpha_c$) as well as the localized phase (for $\alpha > \alpha_c$). This feature implies the fulfillment of the scaling relation among $\beta$, $\delta$, and $\gamma$. Indeed, using $(\alpha - \alpha_c)^{-\gamma}\epsilon_{cr} = \epsilon_{cr}^0(\alpha - \alpha_c)^\beta = \epsilon_{cr}^{1/\delta}$, we get the scaling relation

$$\beta = \frac{\gamma}{\delta - 1}. \quad (21)$$

For the SBM, the same scaling relation has been confirmed for the regime $0 < s < 1/2$ [23]. The ED results are shown for several $\alpha < \alpha_c$. Using $\eta = 10^{-7}$ and $N_b = 3500$, we are able to reproduce the results of adiabatic limit at $\eta = 0$ only for small $\alpha$ values. For larger $\alpha$, the error due to finite $\eta$ and $N_b$ becomes large, and the exact results for $\eta = 0$ and $N_b = \infty$ are more and more difficult to obtain using ED.

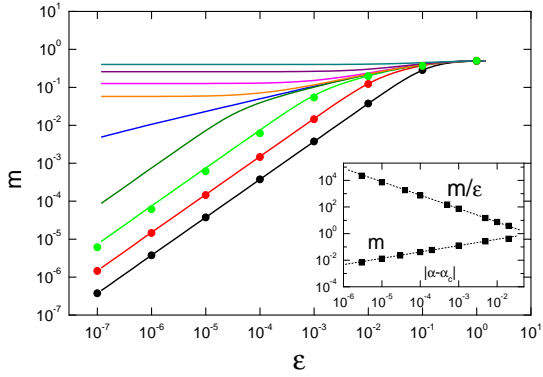

**Fig. 3.** Order parameter $m \equiv \langle\sigma_z\rangle/2$ at $T = 0$ as functions of $\varepsilon$ for various $\alpha$'s. The solid lines are results of adiabatic approximation. Solids dots are ED results using $\eta = 10^{-7}$ and $N_{b0} = 3500$. From top to bottom, $\alpha = 0.05, 0.035, 0.031, 0.0302, 0.03, 0.0299, 0.029, 0.025, 0.01$. Among them $\alpha_c = 0.03$ has slope $1/3$. Other parameters are $s = 0.3$, $\Delta = 0.2$, and $\Lambda = 2.0$. Inset: $m/\epsilon$ (for $\alpha < \alpha_c$) and $m$ (for $\alpha > \alpha_c$) in the small $\epsilon$ limit, as functions of $|\alpha - \alpha_c|$. The dashed lines have slopes of $1/2$ (for $m$) and $-1$ (for $m/\epsilon$).

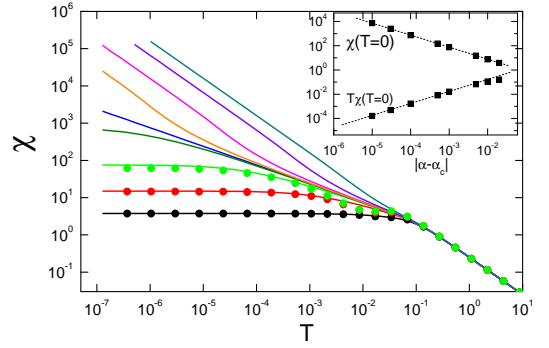

**Fig. 4.** Magnetic susceptibility $\chi \equiv -\partial(\langle\sigma_z\rangle/2)/\partial\epsilon|_{\epsilon=0}$ as functions of $T$ for various $\alpha$'s. The solid lines are results of adiabatic approximation. Solid dots are ED results. From top to bottom, $\alpha$ values are same as in Fig.3. For $\alpha = 0.03$, the slope is $-1/2$. Other parameters are $s = 0.3$, $\Delta = 0.2$, $\varepsilon = 0$, and $\Lambda = 2.0$. Inset: $\chi$ (for $\alpha < \alpha_c$) and $T\chi$ (for $\alpha > \alpha_c$) in the small $T$ limit, as functions of $|\alpha - \alpha_c|$. The dashed lines have slopes of $-1$ (for $\chi$) and $1$ (for $T\chi$).

### 3.2 magnetic susceptibility $\chi$

The magnetic susceptibility $\chi(T)$ from the adiabatic approximation reads

$$\chi = \frac{A}{2Z} - \beta\left(\frac{B}{2Z}\right)^2,$$
$$A = \int_{-\infty}^{\infty}\left[\left(1 - \frac{\Delta^2}{u^2}\right)\beta\cosh\left(\frac{\beta u}{2}\right) + \frac{2\Delta^2}{u^3}\sinh\left(\frac{\beta u}{2}\right)\right]e^{-\frac{\beta}{4}cq^2}dq,$$
$$B = 2\int_{-\infty}^{\infty}\frac{\varepsilon + q}{u}\sinh\left(\frac{\beta u}{2}\right)e^{-\frac{\beta}{4}cq^2}dq. \quad (22)$$

Here $u = \sqrt{\Delta^2 + (\varepsilon + q)^2}$. The partition function $Z$ is given by

$$Z = 2\int_{-\infty}^{\infty}\cosh\left(\frac{\beta u}{2}\right)e^{-\frac{\beta}{4}cq^2}dq. \quad (23)$$

In Fig.4, $\chi(T)$ curves are plotted for different $\alpha$ values. We find similar structure as in Fig.3. The ED results using $\eta = 10^{-7}$ and $N_b = 3500$ agree with the adiabatic approximation in the low and high temperature regimes for small $\alpha$. For larger $\alpha$'s and in the crossover regime, obvious deviations occur due to finite $\eta$ and $N_b$. In the low temperature limit, for $\alpha < \alpha_c$, $\chi \propto (\alpha_c - \alpha)^{-\gamma}T^0$. Numerical fit confirms $\gamma = 1$ (inset of Fig.4). At $\alpha = \alpha_c$, it is found that $\chi \propto T^{-x}$ with the exponent $x = 1/2$. This is again same as the expected behavior for the SBM in the regime $0 < s < 1/2$. The BNRG calculation for the SBM gives $x = s$ in this regime, due to the mass flow error. This error has been discussed and partly remedied by a modified BNRG method [18]. For $\alpha > \alpha_c$, $\chi \propto (\alpha - \alpha_c)T^{-1}$ is consistent with $\chi \propto \langle\sigma_z\rangle^2/T$ in the localized phase and $\beta = 1/2$ (inset of Fig.4).

We denote $T^*$ as the crossover temperature separating the quantum critical regime and the delocalized phase (for $\alpha < \alpha_c$) or the localized phase (for $\alpha > \alpha_c$). Similar observations of the scaling form in $\chi(T)$ curves imply the following relations about $T^*$,

$$T^* \propto (\alpha - \alpha_c)^{\gamma/x},$$
$$T^* \propto (\alpha - \alpha_c)^{2\beta/(1-x)}. \quad (24)$$

Since $T^*$ vanishes in the limit $\alpha \to \alpha_c$ as $T^* \propto (\alpha - \alpha_c)^{1/y_t^*}$ [16], the above equations thus connect the critical exponent $\gamma$, $\beta$, $x$ and $y_t^*$ by

$$y_t^* = \frac{x}{\gamma} = \frac{1-x}{2\beta}. \quad (25)$$

Our results in Fig.4 and numerical fittings therefore confirm the classical exponents $y_t^* = 1/2$.

It is noted that the results $\chi(T)$ are different from that of the mean-field approximation, in which the $Z_2$ symmetry of the SBM at $\epsilon = 0$ is broken artificially. At $\alpha = \alpha_c$, the mean-field approximation gives an exponential form for $\chi(T)$, due to the finite spin gap induced by $\Delta\sigma_x$ term. Here, although the spin gap is finite for a fixed adiabatic mode $q$, the integral over $q$ effectively reduce the gap to zero at $\alpha = \alpha_c$, leading to a power law in $\chi(T)$.

### 3.3 spin-spin correlation function $C(\omega)$

For the spin-spin correlation function $C(\omega)$ defined below,

$$C(t) \equiv \frac{1}{2}\langle\{\sigma_z(t), \sigma_z(0)\}\rangle_H,$$
$$C(\omega) = \frac{1}{2\pi}\int_{-\infty}^{\infty}e^{i\omega t}C(t)dt, \quad (26)$$

its expression in the adiabatic limit is obtained as

$$C(\omega) = w_0\delta(\omega) + C_+(\omega)\theta(\omega - \Delta) + C_-(\omega)\theta(-\omega - \Delta). \quad (27)$$

Here

$$C_+(\omega) = \frac{1}{Z}\frac{\Delta^2}{\sqrt{2}|\omega|\sqrt{\omega^2 - \Delta^2}}\left(e^{-\frac{\beta c}{4}q_1^2} + e^{-\frac{\beta c}{4}q_2^2}\right)\cosh\left(\frac{\beta\omega}{2}\right). \quad (28)$$

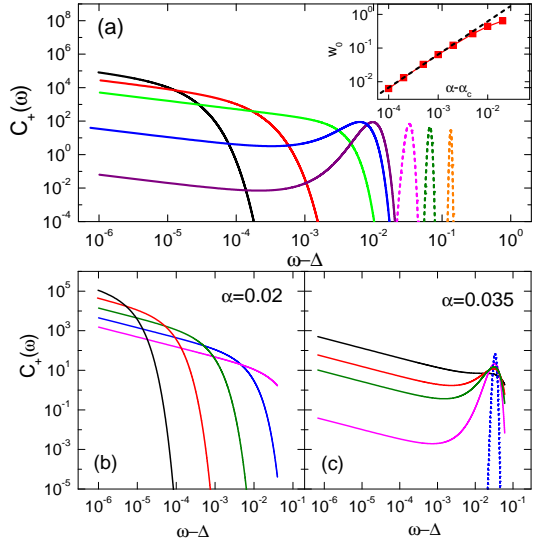

**Fig. 5.** (a) $C_+(\omega)$ versus $\omega - \Delta$ for different $\alpha$'s at $T = 10^{-5}$. From top to bottom on the left: $\alpha = 0.01, 0.025, 0.03, 0.031$, and $0.0315$. The dashed lines are for $\alpha = 0.35, 0.4$, and $0.5$ from left to right, respectively. Inset: $w_0$ (the weight of $\delta(\omega)$) versus $\alpha - \alpha_c$ at $T = 10^{-7}$. The dashed line has a slope 1.0. (b) $C_+(\omega)$ versus $\omega - \Delta$ for $\alpha = 0.02 < \alpha_c$ at different $T$'s. From top to bottom on the left: $T = 10^{-6}, 10^{-5}, 10^{-4}, 10^{-3}, 10^{-2}$. (c) $C_+(\omega)$ versus $\omega - \Delta$ for $\alpha = 0.035 > \alpha_c$ at various $T$'s. From top to bottom on the left: $T = 10^{-3}, 3 \times 10^{-4}, 2 \times 10^{-4}, 10^{-4}$. The dashed line is for $T = 10^{-5}$. Other parameters are $S = 0.3, \Delta = 0.2, \varepsilon = 0, \Lambda = 2.0$.

and $C_-(\omega) = C_+(-\omega)$. The function $\theta(\omega - \Delta)$ is the step function. $q_1$ and $q_2$ are given by

$$q_1 = \sqrt{\omega^2 - \Delta^2} - \varepsilon,$$
$$q_2 = -\sqrt{\omega^2 - \Delta^2} - \varepsilon. \quad (29)$$

It is seen that $C(\omega)$ has a gap at $|\omega| < \Delta$ except a delta peak at $\omega = 0$. The weight of the delta peak $w_0$ is

$$w_0 = \frac{2}{Z} \int_{-\infty}^{\infty} \cosh\left(\frac{\beta u}{2}\right) \frac{(\varepsilon + q)^2}{u^2} e^{-\frac{\beta}{4} cq^2} dq. \quad (30)$$

Here $u = \sqrt{\Delta^2 + (\varepsilon + q)^2}$. From the definition, the sum rule $\int_{-\infty}^{\infty} C(\omega)d\omega = 1$ should hold. Using numerical integral, we have checked that the sum rule is fulfilled by Eq.(27)-(30).

The evolutions of $C_+(\omega)$ with $\alpha$ or $T$ are shown in Fig.5. The weight $w_0$ is finite at $T > 0$. At $T = 0$, it scales like $\langle \sigma_z \rangle^2$: $w_0 \propto (\alpha - \alpha_c)^{2\beta}$ for $\alpha > \alpha_c$ and $w_0 = 0$ for $\alpha < \alpha_c$, as shown in the inset of Fig.5(a). At finite $T$, for $\alpha < \alpha_c$, $C_+(\omega)$ divergence at the gap edge $\omega = \Delta$ as $(\omega - \Delta)^{-1/2}$. As $\alpha$ increases, this peak broadened until at $\alpha = \alpha_c$, a Gaussian peak emerges. As $\alpha$ increases further, the weight continuously shifts from the gap edge to the Gaussian peak centered at $\omega = (\alpha/\alpha_c)\Delta$. These are shown in Fig.5(a). For a fixed $\alpha$, when $T$ decreases, the weight of $C_+(\omega)$ is concentrated either to the gap edge $\omega = \Delta$ with square root divergence, or to the Gaussian peak at $\omega = (\alpha/\alpha_c)\Delta$, depending on whether $\alpha < \alpha_c$ or $\alpha > \alpha_c$, as shown in Fig.5(b) and (c). Either peak becomes sharper with decreasing $T$. Eventually, at $T = 0$, it is expected that $C_+(\omega) = 1/2\delta(\omega - \Delta)$ for $\alpha < \alpha_c$, and $C_+(\omega) = [(1 - w_0)/2]\delta(\omega - \alpha/\alpha_c\Delta)$ for $\alpha \geq \alpha_c$.

It is seen that $C(\omega)$ of the single-mode Hamitonian $H_a$ is nontrivial. At $T = 0$, it describes coherent quantum oscillations with frequency $\Delta$ in the delocalized phase, and with frequency $(\alpha/\alpha_c)\Delta$ in the localized phase. At $T > 0$, the temperature dependent incoherent effects are reflected in the broadening of the delta peaks. These behaviors differ dramatically from that of the SBM. For the latter, at $T = 0$, $C(\omega) \propto \omega^s$ in the small $\omega$ limit. At the critical point, $C(\omega)(T = 0) \propto \omega^{-s}$. Using the RWT-based methods, dynamical properties of the SBM have been studied and rich phenomena disclosed [11,12]. Recently, non-equilibrium dynamics was studied using the non-equilibrium BNRG [30]. The finite temperature dynamical properties of the sub-Ohmic spin-boson model is still a challenge to BNRG.

### 3.4 discussions

3.4.1 scaling relation

Combining the information in Fig.3 and Fig.4, we conclude that these results are consistent with the scaling form of the critical part of the free energy $F_{cri}(\tau, \epsilon, T)$ [31] ($\tau \equiv \alpha - \alpha_c$),

$$F_{cri}(\tau, \epsilon, T) = Tf\left(\frac{\tau}{T^{a_\tau/a_T}}, \frac{\epsilon}{T^{a_\epsilon/a_T}}\right). \quad (31)$$

Here $a_\tau$, $a_T$, and $a_\epsilon$ are scaling constants for $\tau$, $T$, and $\epsilon$, respectively [32]. Indeed, Fig.6 demonstrates that Eq.(31) is fulfilled. From Eq.(31), all the critical exponents discussed above can be expressed using the scaling constants as

$$\beta = (a_T - a_\epsilon)/a_\tau, \quad \delta = a_\epsilon/(a_T - a_\epsilon)$$
$$\gamma = (2a_\epsilon - a_T)/a_\tau, \quad x = (2a_\epsilon - a_T)/a_T,$$
$$y_t^* = a_\tau/a_T. \quad (32)$$

The scaling relations in Eq.(21) and (25) can be regarded as consequences of Eq.(31). The critical properties of the single-mode Hamiltonian $H_0(\eta \to 0)$ can be summarized as $a_\tau/a_T = 1/2$ and $a_\epsilon/a_T = 3/4$.

3.4.2 problem in the BNRG crossover temperature $T^*$

Here we discuss the BNRG result of the the crossover temperature $T^*$ for the SBM. As implied in Eq.(31), $T^* \propto \tau^{1/y_t^*}$ [16] is a temperature scale which marks the crossover between two distinct behaviors in $F_{cri}(T)$. In BNRG, $T^*$ is obtained from the crossover point $N_{cr}$ in the flow of the energy levels using $T^* = \Lambda^{-N_{cr}}$ [4,14,15]. It has the universal meaning as in Eq.(31). For the regime $0 < s < 1/2$, the validity of Eq(31) has been confirmed directly by QMC calculation [16] and indirectly by other studies, with the classical exponents $a_\tau/a_T = 1/2$ and $a_\epsilon/a_T = 3/4$. This implies that $T^* \propto \tau^2$ should hold in $0 < s < 1/2$. However, previous NRG studies gives $T^* \propto \tau^{1/s}$. We therefore conclude that besides the known defect of incorrect exponent $\beta$, $\delta$, and $x$ in the regime $0 < s < 1/2$, a more direct defect in previous BNRG is the wrong crossover energy scale in the flow diagram. We have checked that the boson state truncation does not influence the flow diagram [23]. This problem in the crossover temperature $T^*$ is probably connected to the mass flow error [18]. Given that this issue is under active debate [33], it is still a challenge for us to fully understand the problem with $T^*$ and remedy it within BNRG.

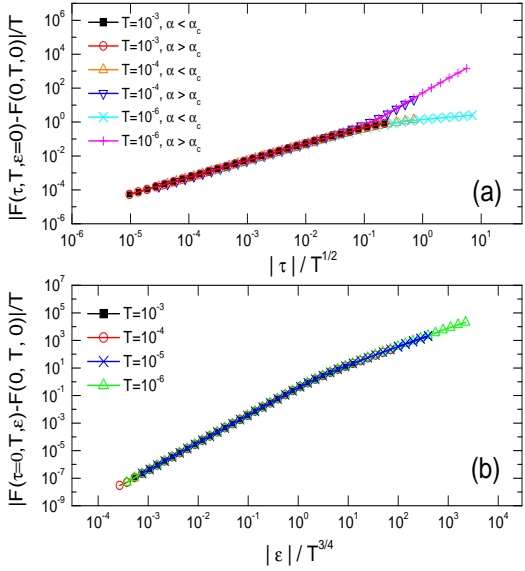

**Fig. 6.** (a) $|F(\tau, T, \epsilon = 0) - F(0, T, 0)|/T$ versus $|\tau|/T^{1/2}$ for various $T$'s in the localized and delocalized phases. (b) $|F(\tau = 0, T, \epsilon) - F(0, T, 0)|/T$ versus $|\epsilon|/T^{3/4}$. Here $\tau \equiv \alpha - \alpha_c$. Other parameters are $s = 0.3$, $\Delta = 0.2$, $\Lambda = 2.0$

The correct result $\nu = 1/s$ for $0 < s < 1/2$ obtained in previous BNRG studies [4,14] is a result of cancelation of errors. It is obtained from the incorrect exponent $y_t^* = s$ fitted from the BNRG data, and the relation $y_t^* = 1/\nu$ which is not appropriate for systems above the upper critical dimension. $\nu = 1/s$ should be correctly obtained using $y_t^* = 1/2$ and $y_t^* = 1/\nu + 1/2 - s$, instead.

### 3.4.3 truncation error in $H_0(\eta \to 0)$

The single-mode Hamiltonian $H_0(\eta \to 0)$ fails to quantitatively produce the correct $\alpha_c$ for the spin-boson model. We would ask why is the truncation approximation invalid for $\tilde{H}_c$, despite $t_0 \to 0$ in the limit $\eta \to 0$. To make this clear, we carry out the same unitary transformation $V$ used for Eq.(13) to the full chain Hamiltonian $\tilde{H}_c$ before it is truncated to $\tilde{H}_0$, $\bar{H}_c \equiv V\tilde{H}_c V^{-1}$. We obtain

$$\bar{H}_c = \frac{\varepsilon}{2}\sigma_z - \frac{\Delta}{2}\sigma_x + \frac{1}{2}\sqrt{\frac{\eta_0}{\pi}}(1 + \frac{\epsilon_0}{\eta})(b_0 + b_0^\dagger)\sigma_z$$
$$+ \sum_{n=0}^{+\infty}[\varepsilon_n b_n^\dagger b_n + t_n(b_n^\dagger b_{n+1} + b_{n+1}^\dagger b_n)]$$
$$+ \frac{t_0}{2\eta}\sqrt{\frac{\eta_0}{\pi}}(b_1 + b_1^\dagger)\sigma_z. \quad (33)$$

In the limit $\eta \to 0$, the coefficient of the last term is nonzero, $t_0/(2\eta)\sqrt{\eta_0/\pi} \to const. \neq 0$. As a result, dropping out the term $t_0(b_0^\dagger b_1 + b_1^\dagger b_0)$ in $\tilde{H}_c$ means that the last term in the above equation is neglected, which is a strong approximation.

$\bar{H}_c$ is obtained from RWT and the subsequent inverse RWT. It should be on the same boson basis as the original Hamiltonian if the two RWT's cancel each other. Indeed, in the limit $\eta \to 0$, all the terms containing $b_0$ or $b_0^\dagger$ disappears and $\sigma_z$ is coupled to $(b_1 + b_1^\dagger)$ instead. The 0-th mode becomes a redundant degrees of freedom and $\bar{H}_c$ recovers the standard NRG chain Hamiltonian. Numerical calculations confirms the following relation between the coefficients,

$$\lim_{\eta \to 0}\left[\frac{t_0}{\eta}\sqrt{\frac{\eta_0}{\pi}}\right] = \lim_{\eta \to \infty}\sqrt{\frac{\eta_0}{\pi}},$$
$$\lim_{\eta \to 0}\epsilon_n = c\lim_{\eta \to \infty}\epsilon_{n-1},$$
$$\lim_{\eta \to 0}t_n = c'\lim_{\eta \to \infty}t_{n-1}. \quad (34)$$

$c$ and $c'$ tend to 1 in the limit $\Lambda \to 1$. Therefore, in the limit $\Lambda \to 1$ where the discretization error disappears, $\bar{H}_c$ return to the chain Hamiltonian $H_c$ exactly.

### 3.4.4 possible NRG algorithm without boson state truncation error

Since the implementation of BNRG based on $\tilde{H}_c$ does not produce physical fixed point, here we propose a possible Wilson chain Hamiltonian which could avoid the boson state truncation error, based on our present understanding. Starting from $\tilde{H}_c$, we use the replacement $q = (b^\dagger + b)/\sqrt{2}$ and $p = i(b^\dagger - b)/\sqrt{2}$. As done in Ref.[27] for analysis of the adiabatic limit, we further carry out rescaling for $p$ and $q$ by defining

$$\tilde{p} = \frac{\sqrt{2}}{\eta}\sqrt{\frac{\eta_0}{\pi}}p, \quad \tilde{q} = \frac{\eta}{\sqrt{2}}\sqrt{\frac{\pi}{\eta_0}}q. \quad (35)$$

In the limit $\eta \to 0$, according to Eq.(11), the coefficients of $\tilde{p}^2$, $(b_1 + b_1^\dagger)\tilde{p}$, and $\tilde{q}\sigma_z$ vanish. The remaining coefficients are nonzero constants. Then one gets a chain Hamiltonian without vanishing parameters,

$$\tilde{H}_c(\eta = 0) = \frac{\varepsilon}{2}\sigma_z - \frac{\Delta}{2}\sigma_x\cosh(\hat{\chi}) - \frac{\Delta}{2}i\sigma_y\sinh(\hat{\chi})$$
$$+ \tilde{\epsilon}_0\tilde{q}^2 + \tilde{t}_0\tilde{q}(b_1 + b_1^\dagger)$$
$$+ \sum_{n=1}^{+\infty}[\varepsilon_n b_n^\dagger b_n + t_n(b_n^\dagger b_{n+1} + b_{n+1}^\dagger b_n)]. \quad (36)$$

Here, $\hat{\chi} = -i\tilde{p}$ and the non-zero coefficients $\tilde{\epsilon}_0$ and $\tilde{t}_0$ are

$$\tilde{\epsilon}_0 = \lim_{\eta \to 0}\left(\frac{\epsilon_0\eta_0}{\pi\eta^2}\right),$$
$$\tilde{t}_0 = \lim_{\eta \to 0}\left(\frac{t_0}{\eta}\sqrt{\frac{\eta_0}{\pi}}\right). \quad (37)$$

The chain coefficients $\epsilon_n = \lim_{\eta=0}\epsilon_n(\eta)$ and $t_n = \lim_{\eta=0}t_n(\eta)$ are well defined quantities due to Eq.(34). $\tilde{H}_c(\eta = 0)$ has the desired features: (i) in the limit $\Delta = 0$, bosons are not displaced and the localized fixed point is correctly described without problem; and (ii) there is no vanishing parameters in the limit $\eta = 0$ as in $\tilde{H}_c$. Therefore, it is a possible candidate for implementing BNRG without boson state truncation error. Further studies in this direction is in progress and the result will be discussed in a separate publication.

## 4 Summary

In this work, we explored the single-mode approximation for the SBM, using the RWT and the NRG-based transformations. Analytical results for the critical coupling strength $\alpha_c$, the magnetic susceptibility $\chi(T)$, and the spin-spin correlation function $C(\omega)$ at finite temperatures are obtained and confirmed by numerical results. We find that the derived single-mode Hamiltonian gives mean-field $\alpha_c$ and the critical exponents are classical, $\beta = 1/2$, $\delta = 3$, $\gamma = 1$, $x = 1/2$, $y_t^* = 1/2$. All these exponents are in agreement with those of the SBM in $0 < s < 1/2$. $C(\omega)$ has nontrivial behavior reflecting coherent oscillation with temperature dependent damping effects due to environment. We discussed the problem in the crossover temperature $T^*$ in the BNRG calculation, and propose a chain Hamiltonian possibly suitable for implementing BNRG without boson state truncation error.

This work is supported by National Program on Key Basic Research Project (973 Program) under grant 2012CB921704, and by the NSFC under grant number 11074302.

## A Appendix

In this appendix, we present details for calculating $\langle\sigma_z\rangle$, $\chi(T)$, and $C(\omega)$ using the adiabatic approximation which is exact in the adiabatic limit $\eta \to 0$ for $\tilde{H}_0(\eta)$.

The matrix for $H_a(q)$ in Eq.(17) under the $\sigma_z$ eigen bases $(|\uparrow\rangle, |\downarrow\rangle)$ reads,

$$\hat{H}_a = \begin{pmatrix} \frac{\varepsilon}{2} + \frac{q}{2} + \frac{c}{4}q^2 & -\frac{\Delta}{2} \\ -\frac{\Delta}{2} & -\frac{\varepsilon}{2} - \frac{q}{2} + \frac{c}{4}q^2 \end{pmatrix}. \quad (38)$$

The eigen energies are

$$E_1 = \frac{c}{4}q^2 - \frac{u}{2},$$
$$E_2 = \frac{c}{4}q^2 + \frac{u}{2}, \quad (39)$$

with $u \equiv \sqrt{\Delta^2 + (\varepsilon + q)^2}$. $E_1 = E_g$ is the ground state energy. The corresponding eigen states are

$$\Psi_1 = \frac{1}{\sqrt{2u}} \begin{pmatrix} \Delta[u + \epsilon + q]^{-1/2} \\ [u + \epsilon + q]^{1/2} \end{pmatrix}, \quad (40)$$

and

$$\Psi_2 = \frac{1}{\sqrt{2u}} \begin{pmatrix} [u + \epsilon + q]^{1/2} \\ -\Delta[u + \epsilon + q]^{-1/2} \end{pmatrix}. \quad (41)$$

The matrix elements of $\sigma_z$ are $\langle\Psi_1|\sigma_z|\Psi_1\rangle = -(\epsilon + q)/u$, $\langle\Psi_2|\sigma_z|\Psi_2\rangle = -\langle\Psi_1|\sigma_z|\Psi_1\rangle$, $\langle\Psi_1|\sigma_z|\Psi_2\rangle = \langle\Psi_2|\sigma_z|\Psi_1\rangle = \Delta/u$. The order parameter $\langle\sigma_z\rangle$ is calculated using Eq.(16). At zero temperature, the expression reduces to

$$\langle\sigma_z\rangle = \langle\Psi_1|\sigma_z|\Psi_1\rangle|_{q=q_0}, \quad (42)$$

with $q_0$ being the coordinate at the minimum of $E_g(q)$. $q_0$ can be solved by $\partial E_g(q)/\partial q = 0$ which leads to Eq.(19).

For susceptibility $\chi(T) \equiv -1/2 \partial\langle\sigma_z\rangle/\partial\epsilon|_{\epsilon=0}$, one gets the expression

$$\chi = -\frac{1}{2Z} \int_{-\infty}^{\infty} dq \frac{\partial}{\partial\epsilon} Tr\left[\sigma_z e^{-\beta H_a(q)}\right]|_{\epsilon=0}$$
$$- \frac{\beta}{4Z^2} \left[\int_{-\infty}^{\infty} dq Tr\left(\sigma_z e^{-\beta H_a(q)}\right)\right]^2|_{\epsilon=0}. \quad (43)$$

Using the matrix elements of $\sigma_z$ and $E_{1,2}$, one gets the Eq.(22) in the main text.

For the spin-spin correlation function $C(\omega)$, we start from the Lehmann expression

$$C(\omega) = \frac{1}{2Z} \int_{-\infty}^{\infty} dq \sum_{m,n} \left(e^{-\beta E_m} + e^{-\beta E_n}\right) |\langle n|\sigma_z|m\rangle|^2 \delta(\omega + E_n - E_m). \quad (44)$$

Using the $\sigma_z$ matrix elements and $E_{1,2}$, one gets

$$C(\omega) = \frac{1}{Z} \int_{-\infty}^{\infty} dq \left[\left(e^{-\beta E_1} + e^{-\beta E_2}\right) \frac{(\epsilon + q)^2}{u^2}\right] \delta(\omega)$$
$$+ \frac{1}{2Z} \int_{-\infty}^{\infty} dq \left(e^{-\beta E_1} + e^{-\beta E_2}\right) \frac{\Delta^2}{u^2} \left[\delta(\omega - u) + \delta(\omega + u)\right]. \quad (45)$$

After the integral over $q$ analytically, we get Eq.(27)-(30) in the main text.

## B Appendix

In this part, we present the numerical methods for solving the single-mode Hamiltonian $\bar{H}_0$ (Eq.(13)) in the adiabatic limit. In the limit $\eta$ tends to zero, $\lambda/\epsilon_0 \sim \eta^{-(1-s)/2} \to \infty$. In this case, the boson occupation basis for $\bar{H}_0$ is not sufficient to produce an accurate result. We therefore carry out RWT to $\bar{H}_0$ and obtain

$$H_0 = \frac{\epsilon}{2}\sigma_z + \epsilon_0 a^\dagger a - \frac{\Delta}{2}\sigma_+ e^{\hat{x}} - \frac{\Delta}{2}\sigma_- e^{-\hat{x}}. \quad (46)$$

Here we use $a = b_0$ and $a^\dagger = b_0^\dagger$ for convenience. $\sigma_\pm = (\sigma_x \pm \sigma_y)/2$. $\hat{x} = \alpha(a^\dagger - a)$ with $\alpha = \lambda/\epsilon_0 = \sqrt{\eta_0/\pi}(1/\epsilon_0 + 1/\eta)$. To diagonalize $H_0$, we need to first write down the matrix for the operator $e^{\pm \hat{x}}$ in the boson occupation basis $\{|m\rangle\}$ ($m = 0, 1, 2, ..., N_b - 1$). Then we construct $H_0$ in the bases $\{|\uparrow\rangle|m\rangle, |\downarrow\rangle|m\rangle\}$ ($m = 0, 1, ..., N_b - 1$).

Note that $e^{\hat{x}}$ is the unitary operator that diagonalized the displaced harmonic oscillator $\hat{h} = a^\dagger a - \alpha(a^\dagger + a)$, i.e., $e^{-\hat{x}}\hat{h}e^{\hat{x}} = a^\dagger a - \alpha^2$. Therefore we have $\langle m|e^{\hat{x}}|n\rangle = \langle m|n\rangle_\alpha$ with $|n\rangle_\alpha$ the n-th eigen state of $\hat{h}$, i.e., $\hat{h}|n\rangle_\alpha = n - \alpha^2$ ($m, n = 0, 1, ...$).

The eigen wave function of displace harmonic oscillator can be calculated using a number of methods. One could diagonalize $\hat{h}$ in the occupation basis directly, or diagonalize $\hat{\chi}$ first and then calculate its exponential. These methods are found to be inaccurate and slow. One could also use the recursive relation as done in Ref. [4], or use the series summation as down in Ref. [6]. These methods are not satisfactory because for $\alpha > 10^1$ and $n > 10^2$, numerical unstability occurs.

To overcome this problem, we resort to a new recursive algorithm for numerically calculating Laguerre polynomials. The eigen wave function of displaced harmonic oscilltor can be expressed by Laguerre polynomial as [34]

$$\langle m|n\rangle_\alpha \equiv \langle m|e^{\alpha(a^\dagger-a)}|n\rangle \quad (m,n=0,1,2,...)$$
$$= e^{-\frac{1}{2}\alpha^2}\sqrt{\frac{n!}{m!}}L_n^{(m-n)}(\alpha^2)\alpha^{m-n}. \quad (47)$$

Here $\alpha^0 = 1$ is assumed even for $\alpha = 0$. $L_n^{(z)}(x)$ is the associated Laguerre polynomial which obey the following recursive relations

$$nL_n^{(z)}(x) = (2n+z-1-x)L_{n-1}^{(z)}(x) - (n+z-1)L_{n-2}^{(z)}(x),$$
$$L_1^{(z)}(x) = -x+z+1,$$
$$L_0^{(z)}(x) = 1. \quad (48)$$

This recursive relation is stable numerically, but $L_n^{(z)}(\alpha^2)$ overflows for $\alpha > 50$ and $n > 500$. At the same time, the prefactor $e^{-1/2\alpha^2}$ vanishes for large $\alpha$, leading to uncontrolled precision.

To solve this problem, we design a self-adapting recursive algorithm to cancel the divergence. We define

$$I_n^{(z)}(x) \equiv L_n^{(z)}(x)\left[P_n^{(z)}(x)\right]^{K_n^{(z)}(x)},$$
$$P_n^z(x) = e^{-\frac{1}{2}x}x^{z/2}\sqrt{\frac{n!}{(n+z)!}}. \quad (49)$$

The $K_n^{(z)}(x)$ is a tunable parameter for each given $x$ and $n$. The recursive relation for $I_n^{(z)}(x)$ reads

$$nI_n^{(z)}(x) = (2n+z-1-x)I_{n-1}^{(z)}(x)\frac{\left[P_n^{(z)}(x)\right]^{K_n^{(z)}(x)}}{\left[P_{n-1}^{(z)}(x)\right]^{K_{n-1}^{(z)}(x)}}$$
$$-(n+z-1)I_{n-2}^{(z)}(x)\frac{\left[P_n^{(z)}(x)\right]^{K_n^{(z)}(x)}}{\left[P_{n-2}^{(z)}(x)\right]^{K_{n-2}^{(z)}(x)}}. \quad (50)$$

The initial conditions are

$$I_1^{(z)}(x) = (-x+z+1)\left[P_1^{(z)}(x)\right]^{K_1^{(z)}(x)},$$
$$I_0^{(z)}(x) = \left[P_0^{(z)}(x)\right]^{K_0^{(z)}(x)}. \quad (51)$$

To do the calculation, initially we set $K_n^{(z)}(x) = 1/|\ln P_n^{(z)}(x)|$. At the recursive calculation for some $(n,z,x)$, if $|I_n^{(z)}(x)| > 10^{200}$ or $|I_n^{(z)}(x)| < 10^{-200}$ occurs, $K_n^{(z)}(x)$ is respectively enlarged or reduced by a fixed factor, say 1.02. Then the newly defined $I_n^{(z)}(x)$ is recalculated. This process iterates until $|I_n^{(z)}(x)|$ is within the required range. Since $P_n^{(z)}(x)$ is small for a large $x$, this iteration is always able to find a suitable $K_n^{(z)}(x)$ that makes $|I_n^{(z)}(x)|$ within $\left[10^{-200}, 10^{200}\right]$, avoiding the upper or lower overflow. Finally, the matrix element of $e^{\hat{\chi}}$ is calculated using the converged $I_n^{(z)}(x)$ and $K_n^{(z)}(x)$ as

$$\langle m|n\rangle_\alpha = I_n^{(m-n)}(\alpha^2)\left[P_n^{(m-n)}(\alpha^2)\right]^{1-K_n^{(m-n)}(\alpha^2)}. \quad (52)$$

Numerical test shows that this method can produce reliable results for $\alpha \sim 10^3$ and $n \sim 10^5$ without problem.